\documentclass[12pt]{article}

\usepackage{pb-diagram}
\usepackage{latexsym}
\usepackage{amsfonts}

\catcode`\@=11
\def\marginnote#1{}

\newcount\hour
\newcount\minute
\newtoks\amorpm
\hour=\time\divide\hour by60 \minute=\time{\multiply\hour by60
\global\advance\minute by-\hour}\edef\standardtime{{\ifnum\hour<12
\global\amorpm={am}%
        \else\global\amorpm={pm}\advance\hour by-12 \fi
        \ifnum\hour=0 \hour=12 \fi
        \number\hour:\ifnum\minute<10
0\fi\number\minute\the\amorpm}}
\edef\militarytime{\number\hour:\ifnum\minute<10 0\fi\number\minute}

\def\draftlabel#1{{\@bsphack\if@filesw {\let\thepage\relax
   \xdef\@gtempa{\write\@auxout{\string
      \newlabel{#1}{{\@currentlabel}{\thepage}}}}}\@gtempa
   \if@nobreak \ifvmode\nobreak\fi\fi\fi\@esphack}
        \gdef\@eqnlabel{#1}}
\def\@eqnlabel{}
\def\@vacuum{}
\def\draftmarginnote#1{\marginpar{\raggedright\scriptsize\tt#1}}
\def\draft{\oddsidemargin -.5truein
        \def\@oddfoot{\sl preliminary draft \hfil
        \rm\thepage\hfil\sl\today\quad\militarytime}
        \let\@evenfoot\@oddfoot \overfullrule 3pt
        \let\label=\draftlabel
        \let\marginnote=\draftmarginnote

\def\@eqnnum{(\theequation)\rlap{\kern\marginparsep\tt\@eqnlabel}%
\global\let\@eqnlabel\@vacuum}  }

\def\numberbysection{\@addtoreset{equation}{section}
        \def\theequation{\thesection.\arabic{equation}}}

\def\underline#1{\relax\ifmmode\@@underline#1\else
 $\@@underline{\hbox{#1}}$\relax\fi}

\catcode`@=12 \relax

\numberbysection

\topmargin by -\headsep

\def\nonu{\nonumber}
\def\br{\begin{eqnarray}}
\def\er{\end{eqnarray}}
\def\be{\begin{equation}}
\def\ee{\end{equation}}

\def\({\left(}
\def\){\right)}
\def\[{\left[}
\def\]{\right]}

\def\a{\alpha}

\def\b{\beta}

\def\d{\delta}

\def\G{\Gamma}

\def\l{\lambda}
\def\L{\Lambda}

\def\o{\over}

\def\p{\phi}
\def\P{\Phi}
\def\pa{\partial}

\def\s{\sigma}

\def\tp0{\Theta_{+}^{(0)}}
\def\tm0{\Theta_{-}^{(0)}}

\def\vp{\varphi}

\def\f#1#2#3 {f^{#1#2}_{#3}}

\def\win1{{\sf w_{1+\infty}}}

\def\Win1{{\sf W_{1+\infty}}}

\def\NPB#1#2#3{{\sl Nucl. Phys.} {\bf B#1} (#2) #3}

\def\TMP#1#2#3{{\sl Theor. and Math. Phys.} {\bf #1} (#2) #3}

\def\TMP#1#2#3{{\sl Theor. Mat. Phys.} {\bf #1} (#2) #3}

\begin{document}


{\large\bf \textsf{ Type-II B\"acklund Transformations via Gauge Transformations}}
\normalsize
\vskip .4in

\begin{center}
{\bf \small \textsf{A R Aguirre\footnote{aleroagu@ift.unesp.br}, T R Araujo\footnote{taraujo@ift.unesp.br}, J F Gomes\footnote{jfg@ift.unesp.br} and A H Zimerman\footnote{zimerman@ift.unesp.br}}}

\par \vskip .2in \noindent
Instituto de F\'{\i}sica Te\'{o}rica-UNESP\\
Rua Dr. Bento Teobaldo Ferraz 271, Bloco II, \\
01140-070, S\~ao Paulo, Brazil\\
\par \vskip .5in

\end{center}
\begin{abstract}

The construction of type II B\"acklund transformation for the sine-Gordon and the Tzitz\'eica-Bullough-Dodd models are obtained from gauge transformation. An infinite number of conserved quantities are constructed  from the defect matrices.  This  guarantees that the introduction of type II defects  for these models does not spoil their integrability.
In particular, modified energy and momentum are derived   and compared with those presented in recent literature.

\end{abstract}


\vskip .4in

\section{Introduction}

Integrable defects  have been introduced some time ago \cite{corrigan1}  from the Lagrangian point of view which, from the conservation of energy and momentum, lead to the derivation of B\"acklund transformations.  Many examples like the  sine-Gordon, mKdV, nonlinear Schr\"odinger, $A^{(1)}_n$-Toda equations, were explicitly developed \cite{corrigan23}, \cite{jump}.  As a characteristic  of these  models only physical fields were present within the formulation and the associated B\"acklund transformations were called type I \cite{corrigan2}. The construction of type-I B\"acklund transformation for the sine-Gordon and for the non-linear Schr\"odinger models through gauge transformations were also employed in \cite{Habi}, \cite{Anastasia} respectively. 

More recently in ref. \cite{corrigan2} an extension of the method employed in \cite{corrigan23}, \cite{jump} was proposed by introducing an auxiliary field and the associated B\"acklund transformation were named type II \cite{corrigan2}.
Examples of type II B\"acklund transformations were explicitly derived in ref. \cite{corrigan2} for the sine-Gordon and for the Tzitz\'eica-Bullough-Dodd models by imposing conservation of the modified  energy and momentum by  a defect. 
It is worth to note that for  the  supersymmetric extensions of sine-Gordon model \cite{ymai},\cite{ymai2} and Thirring models \cite{aguirre},\cite{alexis} the introduction of  auxiliary fields also appeared.     

The purpose of the present  investigation is to derive the type II B\"acklund transformations via gauge transformations.  The choice of such gauge transformations naturally introduces the auxiliary fields. Here we consider the sine-Gordon and  Tzitz\'eica-Bullough-Dodd  models.  In the specific case of the sine-Gordon model, by a particular limiting procedure, we can reduce the type II B\"acklund transformation to the usual \mbox{type I.}

The problem we address consists in relating two distinct solutions of the 
 linear problem  $\Phi^{(1)}$ and $\Phi^{(2)}$ given by 
\br
\pa_{+} \Phi^{(k)} = -A_{+}^{(k)} \Phi^{(k)}, \qquad \pa_{-} \Phi^{(k)} = -A_{-}^{(k)} \Phi^{(k)}, \qquad k=1,2,
\label{1}
\er
where $A_{\pm}^{(k)}$  are  Lie algebra valued functionals of the field $\varphi_k$.   In eq. (\ref{1}) we have introduced the light cone coordinates $x_{\pm} = {{1 \over 2} }(t \pm x)$ with derivatives $\pa_{\pm} = \pa_t \pm \pa_x$, and $ \partial_{+}\partial_{-}=\partial_{t}^{2}- \partial_{x}^{2}$.
Under the gauge transformation 
\br
\pa_{\pm}K = K A_{\pm}^{(1)} - A_{\pm}^{(2)} K, \label{2}
\er
the linear problem (\ref{1})  implies $\Phi^{(2)} = K \Phi^{(1)}$.  This  matrix $K$ is commonly called \emph{defect matrix} and induces relations \mbox{between} two different field configurations $\varphi_1$ and $\varphi_2$, which are expected to correspond to the B\"acklund transformations.

Based upon the gauge transformation (\ref{2}), we  construct   type II B\"acklund transformations connecting two  field configurations for the explicit examples of the sine-Gordon and Tzitz\'eica-Bullough-Dodd models. 

In section 2  we construct  gauge transformations leading to type II B\"acklund transformations for   the sine-Gordon model.  Apart from rederiving the type II B\"acklund transformation of ref. \cite{corrigan2} we obtain a second solution   and observed that, in fact it corresponds to the symmetry $\vp \rightarrow - \vp$ of the sine-Gordon equation.  By a suitable limiting procedure, the system of type II B\"acklund equations can be reduced to type I.  In section 3 we consider the Tzitz\'eica-Bullough-Dodd model and show that the gauge transformation generates a set of equations that decouples into three subsets.  One of which is solved and shown to correspond to the B\"acklund transformation derived  in \cite{borisov}.  
A further change of variables  shows that these equations reduce to those found in \cite{corrigan2}.
 In subsection 4.1 we employ the formalism of ref. \cite{caudrelier} for type II  defect matrix $K$ to construct an infinite number of conserved quantities for the sine-Gordon model in the presence of such defect.  In particular, we give explicit expressions for the defect energy and momentum.  By analysing specific combinations of defect contributions for the conserved quantities we made contact with the defect Lagrangian of ref. \cite{corrigan2}.  In subsection 4.2 we discuss the conservation laws for the Tzitz\'eica-Bullough-Dodd model involving $3 \times 3$ matrices, 
and extend the formalism of ref. \cite{caudrelier} to the Tzitz\'eica-Bullough-Dodd model in the presence of type II defect. We also have an infinite number of conserved quantities for this kind of defect. In particular, we obtain  the defect energy and momentum and compare with the defect Lagrangian of ref. \cite{corrigan2}.

\section{The sine-Gordon model}

For the sine-Gordon model we have
\begin{equation}
{A_{+}}=\left(
\begin{array}{cc}
 {i\o 4}\pa_+\varphi &  {m\o 2}\l e^{{i\vp}\o 2}\\
  -{m\o 2}\l e^{-{i\vp}\o 2}&  -{i\o 4}\pa_+\varphi
\end{array}\right), \qquad {A_{-}}=\left(
\begin{array}{cc}
 -{i\o 4}\pa_-\varphi &  {m\o 2}\l^{-1} e^{-{i\vp}\o 2}\\
  -{m\o 2}\l^{-1} e^{{i\vp}\o 2}&  {i\o 4}\pa_-\varphi
\end{array}\right),
\label{3}
\end{equation}
and  $\l $ is the spectral parameter.  Consider a gauge transformation with the following form for the matrix,
\br
K = \a + \l^{-1} \b + \l^{-2}\gamma .
\label{4}
\er
Let $\vp_1$ and $\vp_2$ be two distinct field configuration. By introducing variables
\br
p= {{\varphi_1 + \varphi_2}\o 2}, \qquad  q= {{\varphi_1 - \varphi_2}\o 2},
\label{5}
\er
we find from (\ref{2}) and (\ref{4}) that equations for the matrices $\a, \b$ and  $\gamma$  can be  grouped into two subsets:

(i) The first one involves $\a_{11}, \a_{22}, \b_{12}, \b_{21}, \gamma_{11}$, $\gamma_{22}$  and leads to
\br
\a_{11} = a_{11} e^{-{{i}\o 2}q},\, \qquad \a_{22} = a_{11} e^{{{i}\o 2}q}, \qquad
\gamma_{11} = c_{11} e^{{{i}\o 2}q}, \qquad \gamma_{22} = c_{11} e^{-{{i}\o 2}q}.
\label{6}
\er
Parametrizing $\b_{21}$ in terms of an auxiliary field $\L$, 
\br
\b_{21} = b_{21} e^{-i\L}e^{{{i}\o 2}p},
\label{7}
\er
we find  for the equations involving $\pa_+ \a_{11}, \pa_- \gamma_{11}, \pa_+\b_{21}$ and $\pa_- \b_{21}$ respectively,
\br
  i\pa_+ q  &=& {{m}\o {2a_{11}}} (b_{21} e^{-i\L} e^{ip} + e^{-{{i}\o 2}p} \b_{12}), \nonumber \\
 i\pa_- q  &=& -{{m}\o {2c_{11}}} ( e^{{{i}\o 2}p} \b_{12} + b_{21} e^{-i\L}), \nonumber \\
i\pa_+ \L &=& -{{mc_{11}}\o {2b_{21}}} e^{-ip} e^{i\L}( e^{iq} -e^{-iq}), \nonumber \\
i\pa_- (\L -p)&=& {{ma_{11}}\o {2b_{21}}} e^{i\L}( e^{iq} -e^{-iq}),
\label{8}
\er
together with
\br
\pa_+ \b_{12} &=& -{{i}\o {2}} \pa_+ p \,\b_{12} + {{mc_{11}}\o 2 }e^{{{i}\o 2}p} (e^{iq} - e^{-iq}), \nonu \\
\pa_- \b_{12} &=& {{i}\o {2}} \pa_- p \,\b_{12} - {{ma_{11}}\o 2 }e^{-{{i}\o 2}p} (e^{iq} - e^{-iq}).
\label{9}
\er
A solution  for (\ref{9}) compatible with (\ref{8}) is found to be
\br
\b_{12} = -{{b_{21}}\o 4} e^{-{{i}\o 2} p +i\L} ( e^{iq} + e^{-iq} +\eta),
\label{10}
\er
where $\eta$ is an arbitrary constant.  Therefore 
\br
 K &=& \left[
   \begin{array}{cc c} 
    \, e^{-\frac{iq}{2} } -\frac{1}{\l^2}c_{11}\,e^{\frac{iq}{2} } & \quad \mbox{}& -\frac{1}{4\l}b_{21} \, e^{i\L} \,e^{-\frac{ip}{2}}(e^{iq} + e^{-iq} + \eta )\\[0.4cm]
     \frac{1}{\l} b_{21}\,e^{-i\L} \,e^{\frac{ip}{2}}    &   \quad \mbox{}& \,e^{\frac{iq}{2}}-\frac{1}{\l^2}c_{11}\,e^{-\frac{iq}{2} }
   \end{array}
 \right], \label{11}
\er
where we have chosen $a_{11}=1$.

  For $c_{11} = 0$ and in  the limiting case where $\L = i\tilde \L $,  with $\tilde \L$ constant,  for large  
   $  \eta, \tilde \L$, and small $b_{21}$ limit we find
\br
b_{21}e^{-i\L} = B, \qquad {b_{21}}e^{i\L} \rightarrow 0, \qquad {{b_{21}}\o {4 }}e^{i\L} \eta  \rightarrow B,
\label{12a}
\er
where $B$ is a finite constant. Then, we obtain
\br
 K &=& \left[
   \begin{array}{cc c} 
    \, e^{-\frac{iq}{2} }\, & \quad \mbox{}& -\frac{1}{\l} B \,e^{\frac{-ip}{2}}\\[0.4cm]
     \frac{1}{\l} B \,e^{\frac{ip}{2}}    &   \quad \mbox{}& \,e^{\frac{iq}{2}}
   \end{array}
 \right].\label{13}
\er
This has the structure of type I B\"acklund transformation \cite{corrigan1}. 
In the limit specified in (\ref{12a}) eqs. (\ref{8}) reduces to 
\br
\pa_- p = -{{m}\o {B}} \sin ( q ), \qquad 
\pa_+ q = m B \sin ( p ).
\label{13l}
\er
Now, by introducing $\s =-{{2}\o {b_{21}}} = {{b_{21}}\o {2c_{11}}}$ and using eq. (\ref{12a}), the equation (\ref{8}) becomes  (with $a_{11}=1$),
\br
 i\pa_-(p-\L) &=& \frac{m\s}{4}\,e^{i\L}(e^{iq} - e^{-iq}),\nonu \\
 i\pa_+\L &=& -\frac{m}{4\s}\,e^{-i(p-\L)}(e^{iq} - e^{-iq}),\nonu \\
 i\pa_- q &=& \frac{m\s}{4}( e^{i\L}(e^{iq} +e^{-iq} + \eta ) - 4 \,e^{-i\L } ), \nonu \\
  i\pa_+ q &=& \frac{m}{4\s}( e^{-i(p-\L)}(e^{iq} +e^{-iq} + \eta ) - 4\, e^{i(p-\L )} ), \label{14}
\er
and the expression for $K$ in eq. (\ref{11}) takes the form
\br
 K &=& \left[
   \begin{array}{cc c} 
    \, e^{-\frac{iq}{2} } -\frac{1}{(\s\l)^2}\,e^{\frac{iq}{2} } & \quad \mbox{}& \frac{1}{2(\s\l)} \, e^{i\L} \,e^{-\frac{ip}{2}}(e^{iq} + e^{-iq} + \eta ) \\[0.4cm]
     -\frac{2}{(\s\l)} \,e^{-i\L} \,e^{\frac{ip}{2}}    &   \quad \mbox{}& \,e^{\frac{iq}{2}}-\frac{1}{(\s\l)^2}\,e^{-\frac{iq}{2} }
   \end{array}
 \right]. \label{17}
\er
By cross-differentiating the last two equations in (\ref{14}),  we find that if the field $\varphi_1$ satisfies the sine-Gordon equation
\br
\pa_t^2 \vp -\pa_x^2 \vp =-m^2 \sin \vp,
\label{sg}
\er
then the field $\varphi_2$ also satisfies it. In addition, differentiating the second equation in (\ref{15}) with respect to $x_-$, we obtain
\br
 \pa_-\pa_+(i\L) = -\frac{m^2}{16}\,e^{-ip}\( 4\,e^{2i\L} - (e^{iq} +e^{-iq}) (4- \eta \,e^{2i\L}) \) .
 \label{15}
\er
Equations (\ref{14}) and (\ref{15}) were considered in \cite{corrigan2}. \\

(ii) The second set of equations  involves $\a_{12}, \a_{21}, \b_{11}, \b_{22}, \gamma_{12}$ and $\gamma_{21}$ which satisfy
\br
\gamma_{12} &=& \bar c_{12} e^{-{{i}\o 2} p}, \qquad \,\,\,\,\,\,\gamma_{21} = -\bar c_{12} e^{{{i}\o 2} p},\nonu \\
\a_{21} &=& -a_{12}e^{-{{i}\o 2} p}, \qquad \a_{12} = a_{12}e^{{{i}\o 2} p} ,\nonu \\[0.1cm]
\pa_+ \a_{12} &=& -{{i}\o 2} \pa_+  p\, \a_{12} +{{m}\o 2} e^{{{i}\o 2} (p+q)}\b_{11} - {{m}\o 2} e^{{{i}\o 2} (p-q)}\b_{22}, \nonu \\
\pa_+ \a_{21} &=& {{i}\o 2} \pa_+  p \,\a_{21} -{{m}\o 2} e^{-{{i}\o 2} (p+q)}\b_{22} + {{m}\o 2} e^{-{{i}\o 2} (p-q)}\b_{11}, \nonu \\
\pa_+ \b_{11} &=& {{i}\o 2} \pa_+ q \,\b_{11} - {{m}\o 2} e^{-{{i}\o 2} (p+q)}\gamma_{12} - {{m}\o 2}e^{{{i}\o 2} (p-q)}\gamma_{21}, \nonu \\
\pa_+ \b_{22} &=& -{{i}\o 2} \pa_+ q\, \b_{22} + {{m}\o 2} e^{{{i}\o 2} (p+q)}\gamma_{21} + {{m}\o 2}e^{-{{i}\o 2} (p-q)}\gamma_{12},
\label{18}
\er
and 
\br
\pa_- \b_{11} &=& -{{i}\o 2} \pa_-  q \,\b_{11} -{{m}\o 2} e^{{{i}\o 2} (p+q)}\a_{12} - {{m}\o 2} e^{-{{i}\o 2} (p-q)}\a_{21}, \nonu \\
\pa_- \gamma_{12} &=& {{i}\o 2} \pa_-  p \,\gamma_{12} +{{m}\o 2} e^{-{{i}\o 2} (p+q)}\b_{11} - {{m}\o 2} e^{-{{i}\o 2} (p-q)}\b_{22}, \nonu \\
\pa_- \gamma_{21} &=& -{{i}\o 2} \pa_-  p \,\gamma_{21} -{{m}\o 2} e^{{{i}\o 2} (p+q)}\b_{22} + {{m}\o 2} e^{{{i}\o 2} (p-q)}\b_{11}, \nonu \\
\pa_- \b_{22} &=& {{i}\o 2} \pa_-  q \,\b_{22} +{{m}\o 2} e^{-{{i}\o 2} (p+q)}\a_{21} + {{m}\o 2} e^{{{i}\o 2} (p-q)}\a_{12}.
\label{19}
\er
We now propose the change of variables (introducing the auxiliary field $\bar \L$)
\br
\b_{22} = b_{22} e^{-i\bar \L - {{i}\o 2} q}.
\label{20}
\er
The equations involving $\pa_- \gamma_{12},  \pa_+ \b_{22},   \pa_- \b_{22}$  and $ \pa_+ \a_{12}$ yields respectively,
\br 
i\pa_+ p &=& -{{m}\o {2a_{12}}}( e^{-i\bar \L -iq} b_{22} - e^{{{i}\o 2}q} \b_{11} ), \\
i\pa_- (\bar \L + q) &=& -{{m}\o 2} {{ a_{12}}\o b_{22}} \,e^{i\bar \L }(e^{ip} - e^{-ip}), \\
i\pa_+ \bar \L  &=& {{m}\o 2}{{ \bar c_{12}}\o b_{22}}\, e^{i\bar \L +iq} (e^{ip} - e^{-ip}),\\
 i\pa_- p &=& {{m}\o {2\bar c_{12}}} (b_{22} e^{-i\bar \L} - e^{-{{i}\o 2}q} \b_{11} ).
\label{21}
\er
A solution of $\b_{11}$ compatible with (\ref{18}) and (\ref{19}) is
\br
\b_{11} = -{{a_{12}\bar c_{12}}\o {b_{22}}} e^{i(\bar \L +{{1}\o 2}q)}(e^{ip} + e^{-ip} +\bar \eta ),
\label{22}
\er
where $\bar \eta$ is an arbitrary constant.   $K$ is then given in the following form
\br
 K &=& \left[
   \begin{array}{cc c} 
    \, -{{1}\o {\l}} {{a_{12} \bar c_{12}}\o {b_{22}}}e^{i(\bar \L + {{1}\o 2}q)}(e^{ip} + e^{-ip} +\bar \eta)
     & \quad \mbox{}& a_{12} e^{{{i}\o 2}p} + {{1}\o {\l^2}} \bar c_{12} e^{-{{i}\o 2}p} \\[0.4cm]
    -a_{12} e^{-{{i}\o 2}p} - {{\bar c_{12}}\o {\l^2}}e^{{{i}\o 2}p}    &   \quad \mbox{}& \,{b_{22}\o {\l}}e^{-i(\bar \L +{{1}\o 2}q)}
   \end{array}
 \right]. \label{23}
\er
In the limit $\bar c_{12} = 0$  with large real constants $i\bar \L, \bar \eta$ and small $b_{22}$  satisfying 
\br
b_{22} e^{-i\bar \L} = \bar B, \qquad -{{a_{12}\bar c_{12}}\o {b_{22}}}e^{i\bar \L}\rightarrow 0, \qquad -{{a_{12}\bar c_{12}}\o {b_{22}}}e^{i\bar \L} \bar \eta = \bar B,
\label{24}
\er 
where $\bar B$ is a constant, we recover  the structure of type I B\"acklund transformations,
\br
\pa_+ p = m{{\bar B}\o {\bar a_{12}}} \sin (q), \qquad \pa_- q =-m {{\bar a_{12}}\o {\bar B}} \sin (p).
\label{25}
\er
Notice that (\ref{25}) can be obtained from (\ref{14}) by exchanging $\vp_2 \rightarrow - \vp_2$.

\section{The Tzitz\'{e}ica-Bullough-Dodd model}

 The Tzitz\'{e}ica-Bullough-Dodd  model is given by the field equation
\begin{equation}
\partial_{+} \partial_{-}\phi=-e^{\phi}+e^{-2\phi}. \label{ttequation}
\end{equation}
This can be derived from the zero curvature condition or Lax-Zakharov-Shabat equation,
\begin{equation}
\partial_{+}A_{-}-\partial_{-}A_{+}+\left[A_{+},A_{-}\right]=0,\label{zerocurv}
\end{equation}
where $A_{\pm}$  are given by \cite{luis07}:
\begin{equation}
{A_{+}}=\left(
\begin{array}{ccc}
 0 & -i\,\lambda e^{\phi} & 0 \\
 0 & 0 & -i\,\lambda e^{\phi}  \\
 \lambda e^{-2\,\phi}  & 0 & 0
\end{array}\right), \qquad
{A_{-}}=\left(
\begin{array}{ccc}
 -\partial_{-} \phi &\phantom{x}0\phantom{x} & -\frac{1}{\lambda} \\
 -\frac{i}{\lambda} & \phantom{x}0\phantom{x} & 0  \\
 0 & -\frac{i}{\lambda} & \partial_{-} \phi 
\end{array}\right),  \\
\end{equation}
with $\lambda$ being the spectral parameter.  Now, by redefining  $v=e^{\phi}$ \cite{borisov},  the field equation (\ref{ttequation}) becomes:
\br
\pa_+\pa_- v=\frac{\partial_{+}v \, \partial_{-}v}{v}-v^{2}+\frac{1}{v}. \label{tzit}
\er

Let us  now assume the form for the matrix $K$ (\ref{2}), following ref. \cite{borisov} as follows, 
\br
K = \a + {{1}\o \l} \b + {{1}\o \l^2}\delta + {{1}\o \l^3} \gamma ,
\label{t}
\er
where $\a, \b, \d$ and $\gamma$ are $3\times 3$ matrices.
Equation (\ref{2}) decomposes into three independent systems of equations.  We will consider the one  involving  variables $ \{ \a_{11}, \a_{22}$, $\a_{33}, \b_{13}$, $\b_{21}, \b_{32}$, $\d_{12}, \d_{23}, \d_{31}$, $\gamma_{11}, 
\gamma_{22}, \gamma_{33}\} $ such that
\br
{K}=\left(
\begin{array}{ccc}
 \alpha_{11}+ \frac{1}{\lambda^{3}}\gamma_{11} & \frac{1}{\lambda^{2}}\d_{12} & \frac{1}{\lambda^{}}\b_{13} \\[0.2cm]
 \frac{1}{\lambda^{}}\b_{21} & \alpha_{22}+ \frac{1}{\lambda^{3}}\gamma_{22}  & \frac{1}{\lambda^{2}}\d_{23} \\[0.2cm]
  \frac{1}{\lambda^{2}}\d_{31}& \frac{1}{\lambda^{}}\b_{32} & \alpha_{33}+ \frac{1}{\lambda^{3}}\gamma_{33} 
\end{array}
\right).
\label{TT}
\er
Equations (\ref{2}) with $K$ given above are satisfied for 
\br
\gamma_{11} &=& \gamma_{22} \,\,=\,\,\gamma_{33} \,\,=\,\,\nu \,\,=\,\, const, \nonu \\
\a_{11} &=& \xi{{v_2}\o v_1} \,\,=\,\, \a, \qquad \a_{22} = \xi = const, \qquad \a_{33} = \xi{{v_1}\o v_2}= {{\xi^2}\o \a}.
\er
Introducing new variables
\br
\b_{21} = {{Y}\o v_1} = {{\a Y}\o \xi v_2},
\er
the matrix K given in (\ref{TT}) leads to
\br
\mathbf{K}=\left(
\begin{array}{c c c}
 \alpha+\nu \,\lambda^{-3}   & \quad \frac{2\,\xi\, {v_2}\, \nu}{\alpha Y}\left(\alpha+\xi\right)\,\lambda^{-2} & \quad
 \frac{2\,\xi^{2}{v_2}^{2}\,\nu}{\alpha^{2}Y^{2}}\left(\alpha+\xi\, \right)^{2}\lambda^{-1} \\[0.2cm]
 \frac{\alpha Y}{\xi {v_2}}\lambda^{-1} & \xi+ \nu \,\lambda^{-3} & \frac{2\,\xi\, {v_2}\,\nu}{\alpha Y}\left(\alpha+\xi\right)\,\lambda^{-2} \\[0.2cm]
 \frac{\alpha \, Y^{2}}{2 {v_2}^{2} \xi^{2}}\,\lambda^{-2} & \frac{Y}{{v_2}}\lambda^{-1} & \xi^{2}\,\frac{1}{\alpha}+\nu\,\lambda^{-3}
\end{array}
\right),
\label{kk}
\er
The fields $\a$ and $Y$ satisfy the following equations
\br
& \partial_{+}\alpha-i\,\frac{\alpha\,Y}{\xi}-\frac{2\, \nu}{Y^{2}}\left(\alpha+\xi\,\right)^{2}=0 \label{alpha+}, \\[0.1cm] 
& \frac{1}{{v_2}}\partial_{+}Y-\frac{Y}{{v_2}^{2}}\partial_{+}{v_2}+\frac{2\, \nu\,\xi}{\alpha\,{v_2}\,Y}\left(\alpha+\xi\,\right)
+i\,\frac{Y^{2}}{2\,{v_2}\,\xi}=0,\\[0.1cm] 
&\partial_{-}\alpha-i\,\frac{2\,\nu\,{v_2}\,\xi}{Y}\left(\alpha+\xi\,\right)+\frac{\alpha^{2}Y^{2}}{2\,{v_2}^{2}\,\xi^{2}}=0, \\ [0.1cm] 
& \frac{\partial_{-}Y}{{v_2}}+i\,\xi\left(\frac{\xi}{\alpha}-1\right)=0, \label{alpha-}
\er
which were written in \cite{borisov}. We now define the functions
\br
p=\frac{\phi_1+{\phi_2}}{2}, \qquad q=\frac{\phi_1-{\phi}_2}{2}, \qquad \alpha= \xi \exp\left(-2\,q\right).
\er
It then follows that equations (\ref{alpha+}) to (\ref{alpha-}) are given now by 
\br
\partial_{+}q &=& -\frac{1}{2}\( \frac{i}{\xi}e^{\Lambda}+2\,\nu\,\xi\, e^{-2\Lambda}\left(e^{q}+e^{-q}\right)^{2}\),
 \nonu \\
 \partial_{-}q &=& -\frac{1}{2}\(2i\,\nu\,\xi\,e^{p-\Lambda}\left(e^{q}+e^{-q}\right)-\frac{e^{2\Lambda-2p}}{2\,\xi}\), \nonu \\
\pa_+ (\L -p) &=& -\nu \xi e^{-2\L} (e^{2q} - e^{-2q}), \nonu \\
\pa_- \L &=& i\xi e^{-\L+p} (e^{-q}-e^{q}),
\label{q-}
\er
where  $ Y\equiv e^{\Lambda}$.
These equations correspond to those given in \cite{corrigan2} with $\L \rightarrow -\l, \pa_{\pm} \rightarrow \pa_{\mp}$, and $ \phi \rightarrow -\phi$.

\section{Conservation laws for type II defect matrices }

\subsection{The sine-Gordon model}

The conservation laws  for the sine-Gordon model with defects can be derived \cite{caudrelier} by making use of   the equations of motion as a compatibility condition for the associated linear problem,
\br
 \pa_x \Phi(x,t;\l) &=& U(x,t;\l)\,\Phi(x,t;\l), \nonu \\
 \pa_t \Phi(x,t;\l) &=& V(x,t;\l)\,\Phi(x,t;\l),
 \label{l1}
\er
where the Lax pair is taken in the following form
\br
  U &=&
 \left[
   \begin{array}{cc} -\frac{i}{4}\big( \pa_t \vp\big)   & q(\l)\\[0.3cm]
r(\l)  &   \frac{i}{4}\big(\pa_t \vp\big) 
   \end{array} \right]
 , \qquad 
 V \,=\,
 \left[
   \begin{array}{cc} -\frac{i}{4}\big( \pa_x \vp\big)   & A(\l) \\[0.3cm]
   B(\l) & \frac{i}{4}\big( \pa_x \vp\big) 
   \end{array} \right]
 ,\qquad \mbox{}\label{laxsine}
\er
and we have defined the following fields, 
\br
q(\l)&=& -\frac{m}{4}\big(\l e^{\frac{i\vp}{2}} -\l^{-1}e^{-\frac{i\vp}{2}} \big) , \qquad 
r(\l) \,=\,     \frac{m}{4}\big(\l e^{-\frac{i\vp}{2}} -\l^{-1}e^{\frac{i\vp}{2}} \big),\\
A(\l) &=& -\frac{m}{4}\big(\l e^{\frac{i\vp}{2}} +\l^{-1}e^{-\frac{i\vp}{2}} \big) , \qquad 
B(\l) \,=\, \frac{m}{4}\big(\l e^{-\frac{i\vp}{2}} +\l^{-1}e^{\frac{i\vp}{2}} \big) .
\er
From the linear system (\ref{l1}), we can derive two conservation equations, 
\br
 \pa_t \( q\Gamma -\frac{i}{4}(\pa_t\vp) \) = \pa_x \( A\Gamma -\frac{i}{4}(\pa_x\vp) \) \label{e6.7},\\
 \pa_t \( r\Psi +\frac{i}{4}(\pa_t\vp) \) = \pa_x \( B\Psi +\frac{i}{4}(\pa_x\vp) \) \label{e6.8},
\er
where we have introduced the auxiliary functions $\G=\Phi_2\Phi_1^{-1}$ and $\Psi =\Phi_1\Phi_2^{-1}$. These functions satisfy a set of Ricatti equations,
\br
 \pa_x\G &=& r +\frac{i}{2}(\pa_t\vp) \G - q\G^2,\label{e6.9} \\
  \pa_x\Psi &=& q -\frac{i}{2}(\pa_t\vp) \Psi - r\Psi^2, \label{e6.10} 
\er
Firstly, let us consider the equation (\ref{e6.9}) to solve $\G$. Hence, expanding $\Gamma$ as $\l \to \infty$
\br
 \G = \sum_{n=0}^{\infty} \frac{\G_n}{\l^n},
\er
we get,
\br
 \G_0 &=& i e^{-\frac{i\vp}{2}}, \\
 \G_1 &=& -\frac{i}{m}(\pa_t\vp + \pa_x\vp ) e^{-\frac{i\vp}{2}},\\
 \G_2 &=& e^{-\frac{i\vp}{2}}\(-\frac{2}{m^2}\,\pa_x(\pa_t\vp+\pa_x\vp) +\frac{i}{2m^2}(\pa_t\vp+\pa_x\vp)^2 +\sin\vp \),\\
 \G_3 &=& \frac{2}{im}e^{-\frac{i}{2}\vp}\( -\frac{2}{m^2} \pa_x^2 (\pa_t \vp + \pa_x \vp) + \frac{i}{m^2} (\pa_t \vp + \pa_x \vp )(\pa_x \pa_t \vp + \pa_x^2 \vp)\nonu \right.\\
 &\mbox{}& +\left. \cos \vp (\pa_x \vp ) + \frac{1}{2} (\pa_t \vp + \pa_x \vp ) e^{-i\vp }\) .
\er
Thus, we have a first infinite set of conserved quantities generated from
\br
 { I} =\int_{-\infty}^{\infty} dx\, \(q\G -\frac{i}{4}(\pa_t\vp) \).\label{e6.14.}
\er
Now, if we consider the expansion of $\G$ as $\l\to 0$,
\br
 \G = \sum_{n=0}^{\infty} \,{{\hat\G}_n}{\l^n},
\er
we obtain the following coefficients,
\br
 {\hat\G}_0 &=& i e^{\frac{i\vp}{2}}, \qquad {\hat\G}_1 \,=\, \frac{i}{m}( \pa_t\vp -\pa_x\vp ) e^{\frac{i\vp}{2}},\\
 {\hat\G}_2 &=&  e^{\frac{i\vp}{2}}\( -\frac{2}{m^2}\,\pa_x (\pa_t \vp -\pa_x\vp ) +\frac{i}{2m^2} (\pa_t\vp -\pa_x\vp )^2  - \sin\vp \),\\
 \hat \G_3 &=& \frac{2}{im}e^{\frac{i}{2}\vp}\( -\frac{2}{m^2} \pa_x^2 (\pa_t \vp - \pa_x \vp) + \frac{i}{m^2} (\pa_t \vp - \pa_x \vp )(\pa_x \pa_t \vp - \pa_x^2 \vp)\nonu \right.\\
 &\mbox{}& -\left.\cos \vp (\pa_x \vp ) + \frac{1}{2} (\pa_t \vp - \pa_x \vp ) e^{-i\vp }\) .
\er

Now, let us consider the Ricatti equation (\ref{e6.10}) to solve for $\Psi$. Clearly, using the same scheme we can obtain the first few coefficients for the auxiliary function. The results are listed below,
\br 
\Psi_0 &=& i e^{\frac{i\vp}{2}}, \qquad \Psi_1\,=\, -\frac{i}{m}( \pa_t\vp + \pa_x\vp ) e^{\frac{i\vp}{2}},\\
 \Psi_2 &=& e^{\frac{i\vp}{2}}\(\frac{2}{m^2}\,\pa_x (\pa_t\vp +\pa_x\vp ) +\frac{i}{2m^2} (\pa_t\vp+\pa_x\vp )^2 -\sin\vp \),\\
 \Psi_3 &=& -\frac{2}{im}e^{\frac{i}{2}\vp}\( \frac{2}{m^2} \pa_x^2 (\pa_t \vp + \pa_x \vp) + \frac{i}{m^2} (\pa_t \vp + \pa_x \vp )(\pa_x \pa_t \vp + \pa_x^2 \vp)\nonu \right. \\
 &\mbox{}& \left. -\cos \vp (\pa_x \vp ) - \frac{1}{2} (\pa_t \vp + \pa_x \vp ) e^{i\vp }\),
\er
and
\br
{\hat\Psi}_0 &=& i e^{-\frac{i\vp}{2}}, \qquad {\hat\Psi}_1\,=\, \frac{i}{m}( \pa_t\vp - \pa_x\vp ) e^{-\frac{i\vp}{2}},\\
 {\hat\Psi}_2 &=& e^{-\frac{i\vp}{2}}\( \frac{2}{m^2}\,\pa_x(\pa_t\vp -\pa_x\vp) +\frac{i}{2m^2}(\pa_t\vp-\pa_x\vp)^2 +\sin\vp \), \\
 \hat \Psi_3 &=& \frac{2}{im}e^{-\frac{i}{2}\vp}\( \frac{2}{m^2} \pa_x^2 (\pa_t \vp - \pa_x \vp) + \frac{i}{m^2} (\pa_t \vp - \pa_x \vp )(\pa_x \pa_t \vp - \pa_x^2 \vp)\nonu \right.\\
 &\mbox{}& +\left. \cos \vp (\pa_x \vp ) - \frac{1}{2} (\pa_t \vp - \pa_x \vp ) e^{-i\vp }\) .
\er
Therefore, 
\br
 {\mathbb  I} &=& \int_{-\infty}^{\infty} dx \(\frac{i}{4}(\pa_t \vp) + r\Psi \), \label{e6.24.}
\er
generates an infinite number of conservation laws.

Now, we will discuss how the presence of internal boundary conditions, or more common\-ly called \emph{jump defects}, modify the conserved charges of the sine-Gordon model using the Lax pair approach. First, let us place a defect at $x=0$, and consider the generating functional of infinite charges given by (\ref{e6.14.}) in the presence of a defect as follows,
\br
 { I} =\int_{-\infty}^{0} dx\, \left[q_1\G[\vp_1] -\frac{i}{4}\(\pa_t\vp_1\) \right] +\int_{0}^{\infty} dx\, \left[q_2\G[\vp_2] -\frac{i}{4}\(\pa_t\vp_2\) \right].
\er
and taking the time-derivative, we have 
\br
\frac{ d I}{dt} &=& \left[A_1\,\G[\vp_1] -\frac{i}{4}\(\pa_x\vp_1\) \right] \bigg|_{x=0} - \left[A_2\,\G[\vp_2] -\frac{i}{4}\(\pa_x\vp_2\) \right] \bigg|_{x=0} \,\,= \,\, -\frac{ d I_{{D}}}{dt} ,
\er
where \cite{caudrelier},
\br
 I_{{D}}&=& \ln \bigg[K_{11} + K_{12}\, \G[\vp_1]\bigg]\bigg|_{x=0} . \label{e6.36.}
\er
Now, following the same procedure for the generating function	 (\ref{e6.24.}), one gets
\br
 \frac{d\, \mathbb I}{dt} &=&  \left[B_1\Psi[\vp_1] +\frac{i}{4}\(\pa_x\vp_1\) \right] \bigg|_{x=0} - \left[B_2\Psi[\vp_2]+\frac{i}{4}\(\pa_x\vp_2\) \right] \bigg|_{x=0} \,\,= \,\, -\frac{ d  \mathbb {I}_{{D}}}{dt},
\er
where
\br
 { \mathbb I_{{ D}}} &=& \ln \bigg[K_{21}\Psi[\vp_1] + K_{22}\bigg]\bigg|_{x=0},\label{e6.39.}
\er
Now, taking into account the type II defect matrix $K$ in (\ref{17}) and using the formulas (\ref{e6.36.}) and (\ref{e6.39.}) to compute the respective defect contributions to the modified conserved quantities for this case, we obtain the following results,
\br
 { ( I_{\rm D})}_{-1} &=& \frac{i}{2\s}\, e^{-i(p-\L)}(e^{iq} +e^{-iq} +\eta), \,\,\quad ({{{I}}_{\rm D}})_{+1} \,\,=\,\,-\frac{i\s}{2}\, e^{i\L}(e^{iq} +e^{-iq} +\eta), \qquad \mbox{}\\ 
({\mathbb I}_{\rm D})_{-1}&=& -\frac{2i}{\s}\,e^{i(p-\L)}, \qquad \qquad \qquad \qquad \,\,\,({{\hat{\mathbb I}}_{\rm D}})_{+1} = 2i\s e^{-i\L}.\label{72}
\er
The corresponding type II defect energy and momentum for the sine-Gordon model can be written as,
\br
 E_D  &=& im \( {(I_D)}_{-1} -  \mathbb ({\mathbb I}_{\rm D})_{-1}- (\hat{ I}_D)_{+1} + (\hat {\mathbb I}_D)_{+1} \) \nonumber \\
 &=& -\frac{m}{2\s} ( 4\,e^{i(p-\L)} + \,e^{-i(p-\L)}(e^{iq} +e^{-iq} +\eta) ) -\frac{m\s}{2}(4 \,e^{-i\L} + e^{i\L}(e^{iq} +e^{-iq} +\eta)),  \nonu \\
P_D  &=& im \( (I_D)_{-1} -  (\mathbb I_D)_{-1}+( \hat{ I}_D)_{+1} - (\hat {\mathbb I}_D)_{+1} \) \nonumber \\
 &=& -\frac{m}{2\s} ( 4\,e^{i(p-\L)} + \,e^{-i(p-\L)}(e^{iq} +e^{-iq} +\eta) ) +\frac{m\s}{2}(4 \,e^{-i\L} + e^{i\L}(e^{iq} +e^{-iq} +\eta)).\nonu \\
 \label{77}
\er
Notice that ${ ( I_{\rm D})}_{-1} - ({\mathbb I}_{\rm D})_{-1} $ and $ ({{{I}}_{\rm D}})_{+1} - ({{\hat{\mathbb I}}_{\rm D}})_{+1}$ corresponds respectively to the quantities  $f$  and $-g$ of ref. \cite{corrigan2}.

Taking into account the type II defect  matrix $K$ in (\ref{23}) and using (\ref{e6.39.}) and (\ref{72}) we obtain the following results for the modified conserved quantities
\br
 { (\tt I_{\rm D})}_{-1} &=& i\frac{\bar c_{12}}{b_{22}}\, e^{i(q+\bar \L)}(e^{ip} +e^{-ip} +\bar \eta )- \frac{1}{m}(\pa_t \vp_1 + \pa_x \vp_1), \nonu\\
 ({\hat{\tt{I}}_{\rm D}})_{+1} &=& i\frac{a_{12}}{b_{22}}\, e^{i\bar \L}(e^{ip} +e^{-ip} +\bar \eta)+\frac{1}{m} (\pa_t \vp_1 -\pa_x \vp_1), \nonu \\ 
({\mathbb I}_{\rm D})_{-1}&=& i\frac{b_{22}}{a_{12}}e^{-i(q+\bar \L)} - \frac{1}{m}(\pa_t \vp_1 +\pa_x \vp_1), \nonu \\
({{\hat{\mathbb I}}_{\rm D}})_{+1} &=& i\frac{b_{22}}{\bar c_{12}} e^{-i\bar \L} + \frac{1}{m} (\pa_t\vp_1 - \pa_x \vp_1),\label{72a}
\er
yielding defect energy and momentum
\br
 E_D  &=& im \( (I_D)_{-1} -  (\mathbb I_D)_{-1} - (\hat{ I}_D)_{+1} + (\hat {\mathbb I}_D)_{+1} \) \nonumber \\
 &=& m\frac{b_{22}}{a_{12}} e^{-i(\bar \L +q)} - m\frac{\bar c_{12}}{b_{22}}e^{i(\bar \L +q)} \( e^{ip} + e^{-ip} + \bar \eta \) - m\frac{b_{22}}{\bar c_{12}}e^{-i\bar \L} + m \frac{a_{12}}{b_{22}}e^{i\bar \L}\(e^{ip}+e^{-ip}+  \bar \eta\),  \nonu \\
P_D  &=& im \( (I_D)_{-1} -  (\mathbb I_D)_{-1}+( \hat{ I}_D)_{+1} - (\hat {\mathbb I}_D)_{+1}) \) \nonumber \\
 &=& m\frac{b_{22}}{a_{12}}e^{-i(\bar \L +q)} - m\frac{\bar c_{12}}{b_{22}}e^{i(\bar \L +q)}\(e^{ip}+e^{-ip} + \bar \eta\) + m\frac{b_{22}}{\bar c_{12}}e^{-i\bar \L} - m \frac{a_{12}}{b_{22}}e^{i\bar \L}\(e^{ip}+e^{-ip} + \bar \eta\).
 \nonu \\
 \label{80}
\er
Notice that   (\ref{77}) and (\ref{80}) reflect the symmetry $\vp_2 \rightarrow -\vp_2$ already pointed  out with respect to eqs. (\ref{25}) and (\ref{13l}). 

\subsection{The Tzitz\'eica-Bullough-Dodd Model}

For this model,  the matrices $U$ and $V,$ are conveniently written as,
\br
 U &=& \left[\begin{array}{ccc} 
                        - \frac{\(\pa_- v\)}{2v} &  \frac{i\l v}{2} & -\frac{1}{2\l} \\[0.2cm] 
                          -\frac{i}{2\l} & 0 &   \frac{i\l v}{2}\\[0.2cm]
                           -\frac{\l}{2} v^{-2} & -\frac{i}{2\l} & \frac{\(\pa_- v\)}{2v}
                       \end{array}
              \right], \qquad 
V \,=\, \left[\begin{array}{ccc} 
                        \frac{\(\pa_- v\)}{2v} &  \frac{i\l v}{2} & \frac{1}{2\l} \\[0.2cm] 
                          \frac{i}{2\l} & 0 &  \frac{i\l v}{2}\\[0.2cm]
                           -\frac{\l}{2} v^{-2} & \frac{i}{2\l} & -\frac{\(\pa_- v\)}{2v}
                       \end{array}
              \right].     \quad\mbox{}      \label{e1.7}
\er
In terms of these, we can write down the set of linear differential equations (\ref{l1}) as,
\br
 \pa_t \Phi_1 &=& \(\frac{\pa_- v}{2v}\) \Phi_1 + \(\frac{i\l v}{2}\) \Phi_2 + \(\frac{1}{2\l}\) \Phi_3, \label{e1.9}\\
 \pa_t\Phi_2 &=& \(\frac{i}{2\l}\) \Phi_1 + \(\frac{i\l v}{2}\) \Phi_3, \label{e1.10}\\[0.1cm]
 \pa_t\Phi_3 &=&  -\(\frac{\l}{2v^2} \) \Phi_1 + \(\frac{i}{2\l}\) \Phi_2 - \(\frac{\pa_- v}{2v}\) \Phi_3, \label{e1.11}
\er
and
\br
 \pa_x \Phi_1 &=& -\(\frac{\pa_- v}{2v}\) \Phi_1 +\(\frac{i\l v}{2}\)  \Phi_2 - \(\frac{1}{2\l}\) \Phi_3, \label{e1.12}\\
 \pa_x \Phi_2 &=& -\(\frac{i}{2\l}\) \Phi_1 +\(\frac{i\l v}{2}\)  \Phi_3, \label{e1.13}\\[0.1cm]
 \pa_x\Phi_3 &=&  -\(\frac{\l}{2v^2} \)  \Phi_1 - \(\frac{i}{2\l}\) \Phi_2  + \(\frac{\pa_- v}{2v}\)\Phi_3.\label{e1.14}
\er
Now, by defining the auxiliary functions $\G_{12} = \Phi_2\Phi_1^{-1}$ and $\G_{13} = \Phi_3\Phi_1^{-1}$, we can construct an infinite set of conservation laws from the equations (\ref{e1.9})  and (\ref{e1.12}) as follows, 
\br
 \pa_t\left(-\frac{(\pa_- v)}{2v} + \(\frac{i\l v}{2}\) \G_{12} -\frac{1}{2\l}\G_{13}\right) &=& \pa_x\left(\frac{(\pa_- v)}{2v} + \(\frac{i\l v}{2}\) \G_{12} +\frac{1}{2\l}\G_{13}\right), \label{e1.15}
\er
where the auxiliary functions satisfy the following coupled Ricatti equations for the $x$-part,
\br
 \pa_x\G_{12} &=&\(\frac{(\pa_- v)}{2v}\)\G_{12} +\(\frac{i\l v}{2}\)\(\G_{13} - (\G_{12})^2\) - \frac{1}{2\l}\(i-\G_{12}\G_{13}\),\\[0.1cm]
 \pa_x\G_{13} &=& \(\frac{(\pa_- v)}{v}\)\G_{13} -\frac{\l}{2} \(\frac{1}{v^2} + iv \,\G_{12} \G_{13}\) - \frac{1}{2\l}\(i\G_{12} - (\G_{13})^2\),
\er
and for the $t$-part,
\br
 \pa_t \G_{12} &=& -\frac{(\pa_- v)}{2v} \G_{12} +\frac{i\l v}{2}\big(\G_{13} - (\G_{12})^2\big) +\frac{1}{2\l}\big(i-\G_{12}\G_{13}\big),\label{e2.40.}\\[0.1cm]
 \pa_t \G_{13} &=& -\frac{(\pa_- v)}{v} \G_{13} -\frac{\l}{2} \(\frac{1}{v^2} + iv \,\G_{12} \G_{13}\) +\frac{1}{2\l}\big(i\G_{12}-(\G_{13})^2\big).\label{e2.41.}
\er
Now,  as usual these differential equations can be recursively solved by considering an expansion in  non-positive powers of $\l$,
\br
 \G_{12} &=& \sum_{n=0}^\infty \frac{\G_{12}^{(n)}}{\l^n}, \qquad  \G_{13} \,=\, \sum_{n=0}^\infty \frac{\G_{13}^{(n)}}{\l^n}.
\er
The lowest coefficients are simply given by,
\br
\G_{12}^{(0)} &=& i(\mu v)^{-1}, \qquad \qquad\,\,\;\G_{13}^{(0)}\,=\, \mu v^{-2}, \\[0.1cm]
\G_{12}^{(1)} &=& -i(\pa_+ v)v^{-2} , \qquad \quad \!\G_{13}^{(1)}\,=\,  \mu^{-1}(\pa_+ v)v^{-3},       \\[0.2cm]         
\G_{12}^{(2)} &=& -\frac{4i\mu }{3} \,\pa_x\left((\pa_+ v) v^{-1}\right) v^{-1} +\frac{i\mu}{3}\left((\pa_+ v) v^{-1}\right)^2 v^{-1} +\frac{2i \mu}{3}\left(1-v^{-3}\right), \\[0.2cm]
\G_{13}^{(2)} &=& \frac{2}{3} \,\pa_x\left((\pa_+ v) v^{-1}\right) v^{-2} -\frac{2}{3}\left((\pa_+ v) v^{-1}\right)^2 v^{-2} -\frac{1}{3}\left(v^{-1}-v^{-4}\right),
\er
where $\mu$ is an arbitrary constant satisfying $\mu^3=-1$. Assuming sufficiently smooth decaying fields as  $| x | \to\pm \infty$ , the corresponding
conserved quantities reads
\br
 I_1 &=& \int_{-\infty}^{\infty}dx\,\left[-\frac{(\pa_- v)}{2v} +\frac{i\l v}{2}\, \G_{12} -\frac{1}{2\l}\G_{13}\right].\label{e2.45}
\er
 By substituting the expansion of the auxiliary functions into above definition, we get an infinite number of conserved charges $I_1^{(k)}$. It is very easy to check that the conserved quantities corresponding to $k=1$ is trivial, and for $k=0$ we obtain a topological term. The first non-vanishing conserved charge is explicitly given by,
\br 
 I_1^{(-1)} &=& \frac{1}{3} \int_{-\infty}^{\infty} dx\, \left[ \frac{1}{2}\big((\pa_+ v) v^{-1}\big)^2 + \(v+\frac{1}{2}v^{-2}\)\right],
\er
where without loss of generality, we have chosen $\mu=-1$. Then, repeating this procedure we can construct another set of conserved quantities corresponding to the expansion of the auxiliary functions in non-negative powers of $\l$, namely,
\br
  \G_{12} &=& \sum_{n=0}^\infty  \hat{\G}_{12}^{(n)}\, \l^n, \qquad  \G_{13} \,=\, \sum_{n=0}^\infty       				\hat{\G}_{13}^{(n)} \, \l^n.
\er
From the Ricatti equations we get,
\br
  \hat{\G}_{12}^{(0)} &=& i\mu , \qquad \,\,\, \hat{\G}_{13}^{(0)}\,=\, \mu^{-1},\\[0.1cm]
  \hat{\G}_{12}^{(1)} &=& 0, \qquad \quad \hat{\G}_{13}^{(1)} \,=\, -(\pa_- v)v^{-1},\\[0.2cm]
  \hat{\G}_{12}^{(2)} &=& -\frac{2i}{3}\,\pa_x\left((\pa_- v)v^{-1}\right) +\frac{i}{3}\left((\pa_- v)v^{-1}\right)^2 -\frac{i}{3}\left(v -v^{-2}\right), \\[0.2cm]
  \hat{\G}_{13}^{(2)} &=&  -\frac{2\mu}{3}\,\pa_x\left((\pa_- v)v^{-1}\right) +\frac{\mu}{3}\left((\pa_- v)v^{-1}\right)^2 -\frac{\mu}{3}\left(v -v^{-2}\right).
\er
From these results and chosing $\mu=-1$, the first non-vanishing conserved charge is given by
\br
 {\hat I}_1^{(+1)} &=& \frac{1}{3} \int_{-\infty}^{\infty} dx\, \left[ \frac{1}{2}\big((\pa_- v)v^{-1}\big)^2 + \(v+\frac{1}{2} v^{-2}\)\right].
\er
Then, we clearly can combine $I_1^{(-1)}$ and ${\hat I}_1^{(+1)}$ in order to obtain the usual energy and momentum quantities. However, it is not enough because we are not considering all the information coming from the Lax pair. So, it is also possible to construct another infinite sets of conserved quantities by considering two more conservation equations  that can be derived from the equations (\ref{e1.10}), (\ref{e1.11}), (\ref{e1.13}) and  (\ref{e1.14}), namely,
\br
 \pa_t\left(- \(\frac{i}{2\l}\) \G_{21} +\(\frac{i\l v}{2}\)\G_{23}\right) &=& \pa_x\left(\(\frac{i}{2\l}\) \G_{21} +\(\frac{i\l v}{2}\)\G_{23}\right), \label{e2.55}\\[0.1cm]
 \pa_t\left(\(\frac{\pa_- v}{2v}\)- \(\frac{\l}{2v^2}\) \G_{31} -\(\frac{i}{2\l}\)\G_{32}\right) &=& \pa_x\left(-\(\frac{\pa_- v}{2v}\)- \(\frac{\l}{2v^2}\) \G_{31} +\(\frac{i}{2\l}\)\G_{32}\right)\!\!,\qquad \,\mbox{}\label{e2.56}
\er
where we have introduced some other auxiliary functions $\G_{21}=\P_1\P_2^{-1}$, $\G_{23}=\P_3\P_2^{-1}$, $\G_{31}=\P_1\P_3^{-1}$, and $\G_{32}=\P_2\P_3^{-1}$. It is quite straightforward that these functions satisfy a set of Ricatti equations that can be written for the $x$-part as follows,
\br
 \pa_x\G_{21} &=& \,\,\,\,\,\(\frac{i\l v}{2}\) - \(\frac{\pa_- v}{2v}\)\G_{21} -  \(\frac{1}{2\l}\)\G_{23} - \(\frac{i\l v}{2}\) (\G_{21}\G_{23}) + \(\frac{i}{2\l}\)(\G_{21})^2,\qquad \mbox{} \\[0.1cm]
 \pa_x\G_{23} &=& -\(\frac{i}{2\l}\) - \(\frac{\l}{2v^2}\)\G_{21} +\(\frac{\pa_- v}{2v}\)\G_{23} + \(\frac{i}{2\l}\) (\G_{21}\G_{23}) - \(\frac{i\l v}{2}\)(\G_{23})^2,\qquad \mbox{} \\[0.1cm]
 \pa_x\G_{31} &=& -\(\frac{1}{2\l}\) - \(\frac{\pa_- v}{2v}\)\G_{31} +  \(\frac{i\l v}{2}\)\G_{32} + \(\frac{i}{2\l}\) (\G_{31}\G_{32}) + \(\frac{\l}{2v^2}\)(\G_{31})^2,\qquad \mbox{} \\[0.1cm]
  \pa_x\G_{32} &=&\,\,\,\,\, \(\frac{i\l v}{2}\)  -  \(\frac{i}{2\l}\) \G_{31} - \(\frac{\pa_- v}{2v}\)\G_{32}+\(\frac{\l}{2v^2}\) (\G_{31}\G_{32}) + \(\frac{i}{2\l}\)(\G_{32})^2,
\er
and for the $t$-part,
\br
 \pa_t\G_{21} &=& \(\frac{i\l v}{2}\) +\(\frac{\pa_- v}{2v}\)\G_{21} + \(\frac{1}{2\l}\)\G_{23} - \(\frac{i\l v}{2}\) (\G_{21}\G_{23}) - \(\frac{i}{2\l}\)(\G_{21})^2,\qquad \mbox{} \\[0.1cm]
 \pa_t\G_{23} &=& \(\frac{i}{2\l}\) - \(\frac{\l}{2v^2}\)\G_{21} -\(\frac{\pa_- v}{2v}\)\G_{23} -\(\frac{i}{2\l}\) (\G_{21}\G_{23}) - \(\frac{i\l v}{2}\)(\G_{23})^2,\qquad \mbox{} \\[0.1cm]
 \pa_t\G_{31} &=& \(\frac{1}{2\l}\) + \(\frac{\pa_- v}{2v}\)\G_{31} +  \(\frac{i\l v}{2}\)\G_{32} - \(\frac{i}{2\l}\) (\G_{31}\G_{32}) + \(\frac{\l}{2v^2}\)(\G_{31})^2,\qquad \mbox{} \\[0.1cm]
 \pa_t\G_{32} &=& \(\frac{i\l v}{2}\)  + \(\frac{i}{2\l}\) \G_{31} + \(\frac{\pa_- v}{2v}\)\G_{32}+\(\frac{\l}{2v^2}\) (\G_{31}\G_{32}) - \(\frac{i}{2\l}\)(\G_{32})^2.
\er
As was already shown, these equations can be recursively solved by introducing an expansion of the respective auxiliary functions in positive and/or negative powers of the spectral parameter $\l$. Doing so, after a lengthy calculation the first few coefficients for these auxi\-liary functions can be determined, and the results are shown in tables \ref{tab 1} and \ref{tab2}.
\begin{table}
\begin{center}
\begin{math}
\begin{array}{| c | c | c | c |}
 \hline \mbox{} & \mbox{} &\mbox{} & \mbox{}  \\[-0.1cm]
 \!\!\!\!\!\!\!\!\!\!\!\!\G_{21}^{(0)} = iv & \quad \mbox{}\G_{23}^{(0)} = -iv^{-1} & \hspace{-1.4cm} {\hat\G}_{21}^{(0)} \,=\, i  &\hspace{-1.2cm} {\hat\G}_{23}^{(0)} \,= -i \\[0.1cm]
 \hline \mbox{} & \mbox{} &\mbox{} & \mbox{}  \\[-0.2cm]
 \!\!\!\!\!\!\!\G_{31}^{(0)} = -v^2 &\!\!\!\mbox{} \G_{32}^{(0)} \,=\, iv &\hspace{-1cm}  {\hat\G}_{31}^{(0)} = -1  &\hspace{-1.5cm} {\hat\G}_{32}^{(0)} \,=\, i \\[0.1cm]
 \hline \mbox{} & \mbox{} &\mbox{} & \mbox{}  \\[-0.2cm]
\quad \,\,\mbox{}\G_{21}^{(1)} \,= -i(\pa_+ v) \quad \mbox{}& \!\!\!\!\!\mbox{} \G_{23}^{(1)}\,=\,0 &\hspace{-1.3cm} {\hat\G}_{21}^{(1)}\, =\, 0  & \quad {\hat\G}_{23}^{(1)} = -i(\pa_- v)v^{-1} \\[0.1cm]
 \hline \mbox{} & \mbox{} &\mbox{} & \mbox{}  \\[-0.2cm] 
  \G_{31}^{(1)} = (\pa_+ v)v & \!\!\!\!\mbox{} \G_{32}^{(1)} \,=\, 0 &\,\,\,\mbox{} {\hat\G}_{31}^{(1)} \,=\, (\pa_- v)v^{-1} \,\mbox{} & \quad \mbox{}{\hat\G}_{32}^{(1)} = -i (\pa_- v)v^{-1} \\[0.1cm] \hline
\end{array}
\end{math}
\end{center}
\caption{The zero-th and first order coefficients.}
\label{tab 1}
\end{table}
\begin{table}
\begin{center}\begin{math}
\begin{array}{| c |}
 \hline \mbox{}  \\[-0.1cm]
  \, \mbox{}\G_{21}^{(2)} \,=-\frac{4}{3}\left[ \pa_x(\pa_+ v) + \frac{1}{2v}\(\pa_- v\)\(\pa_+v\) +\frac{1}{2}\(v^{-1}-v^2\) \right] \quad \mbox{}\\[0.2cm]
 \hline \mbox{}  \\[-0.2cm] 
  \G_{23}^{(2)} \,=\, \frac{2}{3i}\left[ \pa_x(\pa_+ v) v^{-2} + \frac{1}{2v^3}\(\pa_- v\)\(\pa_+v\) +\frac{1}{2}\(v^{-3}-1\) \right] \\[0.2cm]
 \hline \mbox{}  \\[-0.2cm] 
\quad \,\mbox{}  {\hat\G}_{21}^{(2)} \,=\,\frac{2}{3i}\left[ \pa_x(\pa_- v)v^{-1} - \frac{1}{2v^2}\(\pa_- v\)\(\pa_+v\) -\frac{1}{2}\(v^{-2}-v\) \right] \quad \mbox{}\\[0.2cm]
 \hline \mbox{}  \\[-0.2cm] 
\qquad \mbox{} {\hat\G}_{23}^{(2)} \,=\,\frac{4}{3i}\left[ -\pa_x(\pa_- v)v^{-1} + \frac{1}{2v^2}\(\pa_- v\)\(\pa_+v\) +\frac{1}{2}\(v^{-2}-v\) \right] \quad \mbox{}\\[0.2cm]
 \hline \mbox{}  \\[-0.2cm] 
 \!\!\!\!\G_{31}^{(2)} \,=-\frac{2}{3}\left[ \pa_x(\pa_+ v)v + \frac{1}{2}\(\pa_- v\)\(\pa_+v\) +\frac{1}{2}\(1-v^3\) \right] \quad \mbox{}\\[0.2cm]
 \hline \mbox{}  \\[-0.2cm] 
 \!\!\!\!\G_{32}^{(2)} \,=\,\frac{2}{3iv}\left[ \pa_x(\pa_+ v)v + \frac{1}{2}\(\pa_- v\)\(\pa_+v\) +\frac{1}{2}\(1-v^3\) \right] \quad \mbox{} \\[0.2cm]\hline
\end{array}
\end{math}
\end{center}
\caption{Second-order coefficients.}
\label{tab2}
\end{table}

Now, from equations (\ref{e2.55}) and (\ref{e2.56}) we obtain directly the following two generating functions of the conserved quantities,
\br
 I_2(\l) &=& \int_{-\infty}^{\infty} dx \left[ -\(\frac{i}{2\l}\)\G_{21} +\(\frac{iv\l}{2}\)\G_{23} \right], \label{e2charge}\\[0.2cm]
 I_3(\l) &=& \int_{-\infty}^{\infty} dx \left[ \(\frac{\pa_-v}{2v}\) - \(\frac{\l}{2v^2}\)\G_{31} -\(\frac{i}{2\l}\)\G_{32} \right].\label{e3charge}
\er
Then, by substituting the respective expansion of each auxiliary function and using the coe\-fficients showed in tables \ref{tab 1} and \ref{tab2}, we immediately get the first few non-vanishing conserved quantities, which are explicitly given by,
\br 
 I_2^{(-1)} &=&  I_3^{(-1)} \,\,\,=\,\,\,\frac{1}{3} \int_{-\infty}^{\infty} dx\, \left[ \frac{1}{2}\big((\pa_+ v)v^{-1}\big)^2 + \(v+\frac{1}{2} v^{-2}\)\right], \\[0.2cm]
 {\hat I}_2^{(+1)} &=& {\hat I}_3^{(+1)} \,\,\,=\,\,\,\frac{1}{3} \int_{-\infty}^{\infty} dx\, \left[ \frac{1}{2}\big((\pa_- v)v^{-1}\big)^2 + \(v+\frac{1}{2} v^{-2}\)\right].
\er
From the above results, we can notice that there is a simple combination of all these contributions giving us the usual energy and momentum conserved quantities. In fact, if we define the following conseved quantities,
\br
  \mathbb{I}^{(-1)} &=& I_1^{(-1)} + I_2^{(-1)} + I_3^{(-1)}\,=\,\int_{-\infty}^{\infty} dx\, \left[ \frac{1}{2}\big((\pa_+ v)v^{-1}\big)^2 + \(v+\frac{1}{2} v^{-2}\)\right], \\[0.2cm]
  \hat{\mathbb I}^{(+1)} &=& {\hat I}_1^{(+1)} + {\hat I}_2^{(+1)} + {\hat I}_3^{(+1)} \,=\,\int_{-\infty}^{\infty} dx\, \left[ \frac{1}{2}\big((\pa_- v)v^{-1}\big)^2 + \(v+\frac{1}{2} v^{-2}\)\right],
\er
the conserved energy and momentum can be written as
\br
 E &=& \frac{\big(\mathbb{I}^{(-1)}  + \hat{\mathbb I}^{(+1)}\big) }{2}\,\,\,=\,\,\,\int_{-\infty}^{\infty} dx\,\left\{ \frac{1}{2}\left[ \big(\pa_x\p)^2 +(\pa_t\p\big)^2\right] + \(e^{\p}+\frac{1}{2}e^{-2\p}\) \right\},\\[0.2cm]
 P &=&\frac{\big(\mathbb{I}^{(-1)}  - \hat{\mathbb I}^{(+1)} \big)}{2}\,\,\,=\,\,\,\int_{-\infty}^{\infty} dx\, \frac{\(\pa_x v\)\(\pa_t v\)}{v^2} \,=\,\int_{-\infty}^{\infty} dx\, \(\pa_x\p\)\(\pa_t\p\).
\er
Then, it shows that given a Lax pair for the Tzitz\'eica-Bullough-Dodd model we can cons\-truct an infinite set of conserved charges by using all the information coming from the associated linear problem.
%

Now, we will compute the modified conserved charges coming from the defect contributions for the Tzitz\'eica-Bullough-Dodd model using the Lax pair approach. Then, considering the defect placed at $x=0$, the set of infinite charges given by (\ref{e2.45}) in the presence of a defect reads,
\br
 {\mathcal I}_1(\l) &=& \int_{-\infty}^{0}\!dx\left[-\frac{(\pa_- v_1)}{2v_1} +\frac{i\l v_1}{2}\,  \G_{12}(v_1) -\frac{1}{2\l}\G_{13}(v_1)\right] \nonu \\&\mbox{}& \!\!\!+ \int_{0}^{\infty}\!dx\left[-\frac{(\pa_-  v_2)}{2 v_2} +\frac{i\l  v_2}{2}\, {\G}_{12}(v_2) -\frac{1}{2\l}{\G}_{13}(v_2)\right],\qquad \mbox{} 
\er
differentiating with respect to time, we get
\br
 \frac{d \,{\mathcal I}_1(\l)}{dt} &=&\left[\frac{(\pa_- v_1)}{2v_1}+\frac{i\l v_1}{2}\,  {\G}_{12}(v_1) +\frac{1}{2\l}{\G}_{13}(v_1) \right] \bigg|_{x=0} \nonu \\&\mbox{}&\!\!\! - \left[\frac{(\pa_-  v_2)}{2 v_2} +\frac{i\l  v_2}{2}\, {\G}_{12}(v_2) +\frac{1}{2\l}{\G}_{13}(v_2) \right]\bigg|_{x=0}.
\er
Then, using 
the associated linear system we obtain the following relation between the auxiliary functions of each side,
\br
  \G_{12}(v_2) &=& \frac{K_{21} + K_{22}\G_{12}(v_1) + K_{23}\G_{13}(v_1)}{K_{11} + K_{12} \G_{12}(v_1) +K_{13} \G_{13}(v_1)}, \nonu \\[0.2cm]
  \G_{13}(v_2) &=& \frac{K_{31} + K_{32}\G_{12}(v_1) + K_{33}\G_{13}(v_1)}{K_{11} + K_{12} \G_{12}(v_1) +K_{13} \G_{13}(v_1)}.
\er
Now, from the partial differential equations 
satisfied by $K$ and the two Ricatti equations (\ref{e2.40.}) and (\ref{e2.41.}), we finally get that 
\br
 \frac{d \,{\mathcal I}_1(\l)}{dt} = -\frac{d}{dt} \bigg[ \ln \big( K_{11} + K_{12} \G_{12}(v_1) +K_{13} \G_{13}(v_1) \big) \bigg]\bigg|_{x=0}.
\er 
Then, we have that the modified conserved quantities derived from this conservation equation (\ref{e1.15}) is \,\,${\mathcal I}_1(\l) + {\mathcal I}_{D_1}(\l)$, where
\br
 {\mathcal I}_{D_1}(\l) =  \ln \bigg[ K_{11} + K_{12} \G_{12}(v_1) +K_{13} \G_{13}(v_1) \bigg]\bigg|_{x=0}.
\er
From the above formula we can derive two different sets of defect contribution by considering the expansion of the auxiliary functions in positive and negative powers of $\l$. In particular for the $\l^{\pm 1}$-terms, we obtain
\br 
  {\hat{\mathcal I}}_{D_1}^{(+1)} &=& -2i\xi\, e^{(p-\L)}\(e^{q} + e^{-q}\), \qquad \quad  {\mathcal I}_{D_1}^{(-1)} \,\,=\,\, -2\xi \,e^{-2\L}\(e^{q} + e^{-q}\)^2.
\er
As we now know, to obtain the exact form of the corresponding defect contributions to the energy and momentum, we need to consider the others conservation equations and consequently the other charges given in (\ref{e2charge}) and (\ref{e3charge}). Applying the same steps to derive the defect contributions, one gets
\br
 {\mathcal I}_{D_2}(\l) &=&  \ln \bigg[ K_{21} \G_{21}(v_1) + K_{22}  + K_{23} \G_{23}(v_1)\bigg]\bigg|_{x=0}, \\[0.1cm]
 {\mathcal I}_{D_3}(\l) &=&  \ln \bigg[K_{31}\G_{31}(v_1)  + K_{32} \G_{32}(v_1) + K_{33}\bigg]\bigg|_{x=0}.
\er
From these, we find that the first non-vanishing terms are given explicitly by,
\br
  {\hat{\mathcal I}}_{D_2}^{(+1)} &=& -2i \xi e^{(p-\L)}\(e^{q} + e^{-q}\),\, \qquad \quad  {\mathcal I}_{D_2}^{(-1)} \,\,=\,\, \frac{i}{\xi}\,e^{\L},\\[0.1cm]
   {\hat{\mathcal I}}_{D_3}^{(+1)} &=& -\frac{1}{2}\, e^{2(\L-p)}, \qquad \qquad \qquad \qquad {\mathcal I}_{D_3}^{(-1)} \,\,=\,\,\frac{i}{\xi}\,e^{\L}.
\er
Now, defining by analogy the following two conserved quantities,
\br
 {\mathcal I}_D^{(-1)} &=&  {\mathcal I}_{D_1}^{(-1)} +  {\mathcal I}_{D_2}^{(-1)} +  {\mathcal I}_{D_3}^{(-1)}\,\,=\,\,  -2\xi\, e^{-2\L}\(e^{q} + e^{-q}\)^2 +\frac{2i}{\xi}\,e^{\L}, \\[0.1cm]
  {\hat{\mathcal I}}_D^{(+1)} &=&  {\hat{\mathcal I}}_{D_1}^{(-1)} +  {\hat{\mathcal I}}_{D_2}^{(-1)} +  {\hat{\mathcal I}}_{D_3}^{(-1)}\,\,=\,\, -4i \xi e^{(p-\L)}\(e^{q} + e^{-q}\) -\frac{1}{2}\, e^{2(\L-p)},
\er
we can write down the defect energy and momentum as follows,
\br
 E_D&=& \frac{1}{2}\({\mathcal I}_D^{(-1)} +{\hat{\mathcal I}}_D^{(+1)}\) \nonumber \\&=& -\xi\, e^{-2\L}\(e^{q} + e^{-q}\)^2 +\frac{i}{\xi}\,e^{\L}- 2i \xi e^{(p-\L)}\(e^{q} + e^{-q}\) -\frac{1}{4}\, e^{2(\L-p)}, \qquad \mbox{} \\[0.1cm]
 P_D &=&  \frac{1}{2}\({\mathcal I}_D^{(-1)} -{\hat{\mathcal I}}_D^{(+1)}\) \nonumber \\&=& -\xi\, e^{-2\L}\(e^{q} + e^{-q}\)^2 +\frac{i}{\xi}\,e^{\L}+ 2i \xi e^{(p-\L)}\(e^{q} + e^{-q}\) +\frac{1}{4}\, e^{2(\L-p)}.
\er
These are exactly the defect energy and momentum which are obtained by using the Lagrangian formalism. In particular, we can note that $E_D$ corresponds to $f+g$ of ref.  \cite{corrigan2}, which was expected (with the observations given at the end of section 2). 
It is interesting to emphasize that for obtaining the most general form of the defect energy and momentum for the Tzitz\'eica-Bullough-Dodd model is necessary to consider all the contributions co\-ming from all conservation equations and, for both expansions of every each of the auxiliary functions in positive and negative powers of the spectral parameter $\l$.


\section{Conclusions}

In conclusion,  we have constructed the type II B\"acklund transformations via gauge transformations and the corresponding conserved quantities for the sine-Gordon and Tzitz\'eica-Bullough-Dodd models. This approach can be useful to study these models with such type of defects, in particular for issues concerning their integrability. For this purpose, as disscussed in \cite{Habi} and \cite{Anastasia} for the sine-Gordon and non-linear Schr\"odinger models, the involution of the Hamiltonians in the presence of type II defects needs to be clarified.

\vspace{1cm}
\noindent {\bf Acknowledgements}

\noindent ARA acknowledges the financial support of FAPESP. TRA would like to acknowledge the financial support of CAPES. JFG and AHZ would like thank CNPq for financial support. We also thank the referee for helpful suggestions and comments.
\appendix

\vspace{0.5cm}


\begin{thebibliography}{9}

\bibitem{corrigan1}
 P. Bowcock, E. Corrigan and C. Zambon, \emph{Classically integrable field theories with defects}, \mbox{\emph{Int. J. Mod. Phys.}} \textbf{A19S2} (2004) 82 [hep-th/0305022].


\bibitem{corrigan23} P. Bowcock, E. Corrigan and C. Zambon, \emph{Affine Toda field theories with defects}, \emph{JHEP} \textbf{0401} (2004) 056 [hep-th/0401020].

\bibitem{jump} E. Corrigan and C. Zambon, \emph{Jump-defects in the nonlinear Schrodinger model and other non-relativistic field theories}, \emph{Nonlinearity} \textbf{19} (2006)  1447 [nli.SI/0512038].

\bibitem{corrigan2}
 E. Corrigan and C. Zambon, \emph{A new class of integrable defects}, \emph{J. Phys.} {\bf A42} (2009) 475203 [arXiv:0908.3126]. 
 
 
\bibitem{Habi} I. Habibullin and A. Kundu, \emph{Quantum and classical integrable sine-Gordon model with defect}, {\it Nucl. Phys.} B {\bf 795} (2008) 549 [hep-th/0709.4611].
 
\bibitem{Anastasia} J. Avan and A. Doikou, \emph{Liouville integrable defects: the non-linear Schr\"odinger paradigm}, (2011) [hep-th/1110.4728]. 


\bibitem{ymai}
 J.F. Gomes, L.H. Ymai and A.H. Zimerman, \emph{Classical integrable super sinh-Gordon equation with defects}, \emph{J. Phys. A : Math. Gen.} \textbf{39} (2006) 7471 [hep-th/0601014].
 
 \bibitem{ymai2}
 J.F. Gomes, L.H. Ymai and A.H. Zimerman, \emph{Integrability of a classical $N= 2$ super sinh-Gordon model with jump defects}, \emph{JHEP} \textbf{0803:001} (2008) [arxiv:0710.1391].
 
 \bibitem{aguirre}
 A.R. Aguirre, J.F. Gomes, L.H. Ymai and A. H. Zimerman, \emph{Thirring model with jump-defect}, \emph{Proceedings of Science}, \textbf{PoS(ISFTG) 031} (2009)  [nlin.SI:0910.2888v2].
 
 \bibitem{alexis} 
A.R. Aguirre, J.F. Gomes, L.H. Ymai, A.H. Zimerman, \emph{Grassmannian and Bosonic Thirring model with jump defect}, \emph{JHEP} \textbf{1102:017} (2011) [arxiv:1012.1537]. 
 
\bibitem{caudrelier} V. Caudrelier, \emph{On a systematic approach to defects in classical integrable field theories}, \emph{IJGMMP} \textbf{vol.5 No. 7} (2008) 1085-1108 [arXiv:0704.2326].
 
\bibitem{luis07}
 P.E.G. Assis and  L.A. Ferreira, \emph{The Bullough-Dodd model coupled to matter fields}, 
 \NPB{800}{2008}{449} [arXiv:0708.1342]. 
 
 \bibitem{borisov} A.B. Borisov, S.A. Zykov and M.V. Pavlov, \emph{Tzitz\'eica equation and proliferation of nonlinear integrable equations}, \TMP{131}{2002}{550}.
 
 
 \end{thebibliography}
\end{document}